\documentclass[%
 prc,
 amsmath,amssymb,
 aps,
]{revtex4-1}

\usepackage{graphicx}
\usepackage{dcolumn}
\usepackage{bm}
\usepackage{braket}
\usepackage{esvect}
\usepackage{isotope}
\usepackage{booktabs}
\usepackage{array}
\usepackage{fullpage}
\usepackage{times}
\usepackage{fancyhdr,graphicx,amsmath,amssymb}
\usepackage[ruled,vlined]{algorithm2e}
\include{pythonlisting}
\usepackage[utf8]{inputenc}
\usepackage{units}
\usepackage{verbatim}
\usepackage[table]{xcolor}
\usepackage{siunitx}
\usepackage{appendix}

\definecolor{Gray}{gray}{0.9}

\begin{document}


\title{Bootstrapped Block Lanczos for large-dimension eigenvalue problems}
\author{Ryan M. Zbikowski}
\author{Calvin W. Johnson}%
 \email{cjohnson@sdsu.edu}
\affiliation{%
San Diego State University, San Diego, CA, 92117
}%

\date{\today}

\begin{abstract}
The Lanczos algorithm has proven itself to be a valuable matrix eigensolver for problems with large dimensions, up to  hundreds of millions or even tens of billions. The computational cost of using any Lanczos algorithm is dominated by the number of sparse matrix-vector multiplications  until suitable convergence is reached.  Block Lanczos replaces sparse matrix-vector multiplication with sparse matrix-matrix multiplication, which is more efficient, but for a randomly chosen starting block (or pivot), more multiplications are required to reach convergence. We find that a bootstrapped pivot block, that is, an initial block constructed from approximate eigenvectors computed in a truncated space, leads to a dramatically reduced number of multiplications, significantly outperforming both standard vector Lanczos and block Lanczos with a random pivot. A key condition 
for speed-up is that the pivot block have a non-trivial overlap with the final converged vectors. We implement  this approach in a configuration-interaction 
code for nuclear structure, and find 
 a reduction in time-to-solution by a factor of two or more, up to a factor of ten.

\end{abstract}

\maketitle

\section{\label{sec:level1}Introduction}

A common and important numerical task is finding the eigenpairs of Hermitian or real, symmetric matrices~\cite{parlett1980symmetric}.  While the classic Householder algorithm~\cite{GolubMatComp} is the 
go-to general method for finding all eigenpairs, in cases with very large dimensions where one only desires a 
small number of 
extremal eigenpairs, for example low-lying eigenstates of a quantum mechanical Hamiltonian, one 
often turns to the Lanczos and related algorithms~\cite{GolubMatComp}. 
In brief, the method's strength is to use the matrix to 
iteratively construct a new basis for the so-called Krylov subspace; in this subspace the extremal 
eigenpairs converge quickly with the subspace dimension, which can be understood through moments 
of the matrix~\cite{whitehead1978lanczosmoments,whitehead1980moment}.


Lanczos is often prescribed for matrices so large and sparse that direct updating methods like the classic Householder algorithm are computationally impractical.   The conventional Lanczos approach relies on matrix-vector multiplication instead. Although the detailed performance depends upon the system, the lowest eigenpair (i.e., the ground state of a Hamiltonian)
will typically converge to a specified accuracy in a few tens of iterations, and a few hundred iterations 
will often converge the lowest ten or so eigenpairs,  even for very large dimensions.

The  scientific applications of the Lanczos method are vast.  In this paper we use the nuclear 
many-body problem, specifically configuration-interaction in a 
shell-model basis~\cite{BrussardGlaudemans,RevModPhys.77.427}, as our example system, although 
we will make clear our methods and results for people outside this field. See Appendix~\ref{CIcontext} 
for a few more details. 
Lanczos is an ideal method for this problem~\cite{Whitehead}: one is typically interested in the low-lying 
spectra (energy eigenvalues and transition between eigenstates); the basis can easily become very large and 
is, at least in the so-called $M$-scheme, very sparse. 
Many modern configuration-interaction (CI) shell-model codes \cite{caurier:in2p3-00003301,MFDN,johnson2018bigstick,shimizu2013nuclear}  use single-vector Lanczos as an eigensolver on platforms ranging from laptops to 
supercomputers; the basis dimension of the nuclear Hamiltonian matrix can reach $2 \times 10^{10}$\cite{PhysRevC.97.034328}. 

Nonetheless, for very large dimension problems numerical challenges remain, in particular the desire to 
significantly reduce the time-to-solution: on a supercomputer, the difference between one million CPU-hours and 
two million is not trivial. The time-to-solution in turn depends roughly upon two factors: first, how many 
iterations must be carried out to obtain satisfactory convergence, and second, the efficiency of 
each iteration.  Lanczos is frequently applied 
to very sparse matrices, as low as one non-zero matrix element out a million~\cite{Johnson2013}. (Note that 
when the dimensions are $10^{10}$, this still corresponds to terabytes of data.) 
In such a sparse matrix, the indices between non-zero matrix elements, and the corresponding locations on 
initial and final vectors, are often far apart in memory, with the resulting
loss of data locality reduces the effectiveness of the conventional Lanczos approach. As discussed 
below, in our own codes such loss of data locality can affect efficiency by a factor of two or more.
Furthermore, to dramatically reduce  storage one may turn to 
elegant algorithms that reconstruct non-zero matrix elements on-the-fly~\cite{Johnson2013}, saving memory but at the cost of 
additional time.

One approach to mitigate the loss of locality is  the block Lanczos algorithm \cite{GOLUB1977361}, where the matrix-vector multiplication is performed on blocks of vectors; as shown below, this can be constructed to 
explicitly improve data locality as well as amortizing the cost to reconstruct non-zero matrix elements.
Furthermore, 
the convergence proprieties of the block method are often superior to single-vector Lanczos because the method can exploit memory hierarchies\cite{doi:10.1137/0717059,QiaoLiuXu}. The block method has also been shown to have efficient parallel implementation\cite{guarracino2006parallel}.
Recent work has shown block Lanczos to be useful in nuclear structure for precise computations of near-degenerate eigenvalues\cite{SHIMIZU2019372}.


The benchmark results in the work of \cite{SHIMIZU2019372} demonstrate conventional and restarted block Lanczos method can drastically reduce the number of algorithm iterations. This is shown when the starting choice pivot vectors are taken randomly. However, Ref.~\cite{SHIMIZU2019372} also show an example where the convergence the restarted block Lanczos method is further accelerated using a careful construction of the initial pivot block using approximate eigenvectors, which we here call bootstrapping. Bootstrapping can also be used in standard vector Lanczos (P. Navratil, private communication), but only improving the convergence of single eigenvector.

Inspired by previous success in bootstrapping the thick-restart block Lanczos method, here we investigate the non-restarted block Lanczos approach in our nuclear shell model calculations. Our work also seeks to establish criteria by which bootstrapped block Lanczos is effective in reducing the time-to-solution for solving huge sparse symmetric Hamiltonian matrices which describe select many-fermion systems.

We have implemented block Lanczos in the massively parallel CI code {\tt BIGSTICK}, written 
in Fortran 90, which utilizes
a factorization algorithm to reconstruct Hamiltonian matrix elements on-the-fly~\cite{Johnson2013,johnson2018bigstick}. 
In section \ref{sec:level2} we outline each Lanczos method used, and how the configuration-interaction many-body basis is used to create the Hamiltonian matrix.
Section~\ref{sec:level3} shows how the time-to-solution of bootstrapped block Lanczos compares to conventional block Lanczos in two different model systems.
We indeed find  even a modest bootstrapped pivot can lead to significant improvements. We also 
document the intrinsic speed-up of block Lanczos in general.

\section{\label{sec:level2}Methods}

\subsection{The base Lanczos algorithm}

 The Lanczos algorithm works by iteratively constructing an orthonormal basis, forming the so-called Krylov subspace.
 In this basis the transformed matrix  is tridiagonal, and even when truncated the 
 extremal eigenpairs converge rapidly with subspace dimension to the extremal eigenpairs of the full space. Here 
 we provide a brief overview.

Let $\hat{H}$ be some Hermitian or real, symmetric matrix. (In our applications it is the nuclear many-body 
Hamiltonian matrix, but the algorithm is independent of this interpretation.)  Here we use the Dirac notation common in physics, 
where $\hat{a}$ denotes the matrix representation of an operator, $|v \rangle$ a (column) vector, and 
$\langle v |$ an adjoint or row vector. 

Starting with some initial or pivot vector $|v_1\rangle$, one uses $\hat{H}$ to iteratively constructs a sequence of orthonormal vectors: 
\begin{equation}
\begin{aligned}
|v_i\rangle,i&=1,k\quad \langle v_i| v_j \rangle = \delta_{ij}\\
\hat H|v_1\rangle &= \alpha_1|v_1\rangle+\beta_1|v_2\rangle \\
\hat H|v_2\rangle &=\beta_1|v_i\rangle+\alpha_2|v_i\rangle+\beta_2|v_i\rangle \\
\hat H|v_3\rangle &=\quad\quad\quad\beta_2|v_2\rangle+\alpha_3|v_3\rangle+\beta_3 v_4\rangle\\
\hat H|v_i\rangle &=\quad\quad\quad\beta_{i-1}|v_{i-1}\rangle+\alpha_i|v_i\rangle+\beta_i|v_{i+1}\rangle\\
\hat H|v_k\rangle &=\quad\quad\quad\quad\quad\beta_{k-1}|v_{k-1}\rangle+\alpha_k|v_k\rangle
\end{aligned}
\end{equation}
Thus each iteration results in the construction of a new Lanczos vector, 
as well as automatically generating the matrix elements of $\hat{H}$ in this new basis, where
hermiticity leads to a tridiagonal form.
 After $k-1$ iterations one has $k$ Lanczos vectors and a $k$-dimensional Krylov subspace. Through the orthonormality of the Lanczos vectors, one one can show in this basis the Hamiltonian, $\hat{H}$ is tridiagonal:

\begin{equation*}
H^{k} = 
\begin{pmatrix}
\alpha_{1} & \beta_{1} & &  0 \\
\beta_{1} & \alpha_{2} &\ddots &\\
  &\ddots   & \ddots & \beta_k  \\
0 & & \beta_{k-1} & \alpha_{k} 
\end{pmatrix}
\end{equation*}

\begin{algorithm}[h]
$\mathbf{1 -}$ $| w_i \rangle = \hat{H}|v_i \rangle$ Initial matrix-vector product on vector $v_i$;\\
$\mathbf{2 -}$ $\alpha_i = \langle v_i|w_i\rangle$ dot product to recover $\alpha_i$; \\
$\mathbf{3 -}$ $|w_i\rangle \longleftarrow |w_i\rangle -\alpha_{i}|v_i\rangle$ orthogonalize against initial vector $v_i$ \\
$\mathbf{4 -}$ If $i > 1$ $| w_i\rangle \longleftarrow |w_i\rangle -\beta_{i-1}|v_{i-1}\rangle$ orthogonalize against prior vector $v_{i-1}$\\
$\mathbf{5 -}$ $\beta_i = \sqrt{\langle w_i|w_i\rangle}$ calculate norm to get $\beta_i$;\\
$\mathbf{6 -}$ $| v_{i+1} \rangle = \beta^{-1} |w_i\rangle $ Normalize $|w_i\rangle$ to get next Lanczos vector $|v_{i+1}\rangle $
    \caption{{\bf Lanczos method}
    \label{Lanczos method}}

\end{algorithm}
The method described is only correct with perfect arithmetic. Because of round-off error which, if 
unchecked, grows with each iteration, each Lanczos vector must be orthogonalized against all previous Lanczos vectors. 
The cost of reorthogonalization is one of the major weaknesses of the Lanczos algorithm.
Despite this weakness, the Lanczos algorithm is an effective and robust way to find extremal eigenpairs.

The main computational cost of the Lanczos algorithm is the matrix-vector multiplication. This in turn 
depends upon the number of non-zero matrix elements; therefore, representations with small dimensions but 
with higher number of non-zero matrix elements (see, e.g., Table 2 of \cite{dytrych2016efficacy}) may not actually lead to a computational savings. 

At some point, however, the cost of enforced orthogonalization, which goes as the number of previous vectors and 
thus as the number of iterations, can compete with matrix-vector multiplication. This, along with the cost of 
storing all prior vectors, led to the thick-restart Lanczos algorithm~\cite{wu2000thick}, where one keeps only a reduced set of 
vectors: the cost of reorthogonalization and storing Lanczos vectors is reduced, at the price of 
more total iterations. 

There are two other considerations worth regarding. The matrix-vector multiplication is
\begin{equation}
w_i = \sum_j H_{ij} v_j.
\end{equation}
First, when $\hat{H}$ is very sparse, the indices $i,j$ on the final and initial vectors, respectively, 
will generally violate data locality, which in turn leads to cache misses. Empirically
we found in our code that this 
leads to approximately a factor of two in the matrix-vector multiplication time.  Second, in order to 
reduce the storage of the non-zero matrix elements $H_{ij}$, many codes including ours use on-the-fly 
reconstruction of matrix elements. While this reduce the matrix element memory footprint by an order of 
magnitude or more~\cite{Johnson2013}, reconstruction rather than simple look-up in memory also costs time; again, 
our empirical estimates comparing against codes that store non-zero matrix elements is roughly a 
factor of 2-4. In the next section we will show how block Lanczos can mitigate both of these problems.

\subsection{Block Lanczos algorithm}

 Block Lanczos~\cite{GOLUB1977361} is variation of the single vector Lanczos method where the Hamiltonian matrix is applied to a \textit{block} of vectors. The Hamiltonian matrix in the M-scheme basis is very sparse meaning the indices of the accessed vector elements can be far apart in memory leading to a loss of locality. This loss of locality is increased by the complexity of the on-the-fly algorithm.
 
 
 In block Lanczos we apply the Hamiltonian to a block of vectors

\begin{equation}
\begin{aligned}
w_{a,i} = \sum_{j}H_{ij}v_{a,j},
\end{aligned}
\end{equation}
where $a$ labels the different vectors in the blocks. Here we make two important 
choice. First, we make the loop over $a$ the innermost loop: in pseudocode this is
\begin{verbatim}
loop over matrix elements: i,j, H_ij
    loop over a
        w(a,i) = w(a,i) + H_ij * v(a,j)
\end{verbatim}
Then the cost to reconstruct $H_{ij}$ gets amortized over 
the loop over $a$. Furthermore, since Fortran stores arrays column-major, that is, 
for fixed $j$, the array elements $v(a,j)$ and $v(a+1,j)$ are neighbors in memory, 
the loop over $a$ maximizes data locality.  Both of these lead to a significant speed-up, 
by a factor of two or more.

We carry out block Lanczos as follows  
\begin{equation}
\begin{aligned}
\mathbf{HV}_n = \mathbf{W}_n.
\end{aligned}
\end{equation}
The block $\mathbf{V}_n$ is generated iteratively just as standard Lanczos through matrix multiplication. However the transpose of the matrix is stored to improve locality. $\mathbf{W}_n$ is just a temporary matrix. We actually store $\mathbf{W}_n^T$ 
\begin{equation} \label{eqn:WTn}
\begin{aligned}
\mathbf{V}_n^T\mathbf{H} = \mathbf{W}_n^T.
\end{aligned}
\end{equation}

Eq.(\ref{eqn:WTn})  reduces calls to the cache as blocks are stored row-wise in memory. The block Lanczos iteration is
\begin{equation}
\begin{aligned}
\mathbf{H}\mathbf{V}_n = \mathbf{V}_{n-1}\mathbf{B}_{n-1}+\mathbf{V}_{n}\mathbf{A}_{n}+\mathbf{V}_{n+1}\mathbf{B}_{n}.
\end{aligned}
\end{equation}
$\mathbf{A}_{n}$,$\mathbf{B}_{n}$ are $N_\mathrm{block} \times N_\mathrm{block}$  square matrices. The $\mathbf{A}_{n}$ matrices are symmetric, but generally the $\mathbf{B}_{n}$ are not. Next we compute
\begin{equation}
\begin{aligned}
\mathbf{W}_n^T\mathbf{W}_n = \mathbf{A}_n,
\end{aligned}
\end{equation}
and then subtract
\begin{equation}
\begin{aligned}
\mathbf{W}_n-\mathbf{V}_n\mathbf{A}_n = \mathbf{W}_n^{'}= \mathbf{V}_{n+1}\mathbf{B}_n.
\end{aligned}
\end{equation}
We want to extract $\mathbf{V}_{n+1}$, and $\mathbf{B}_n$. To do this we construct the $N_\mathrm{block} \times N_\mathrm{block}$ square symmetric overlap matrix $\mathbf{O}$.
\begin{equation}
\begin{aligned}
\mathbf{O}=\mathbf{W}_n^{'T}\mathbf{W}_n^{'}=\mathbf{B}_n^T\mathbf{V}_{n+1}^T\mathbf{V}_{n+1}\mathbf{B}_n = \mathbf{B}_n^T\mathbf{B}_n
\end{aligned}
\end{equation}
To factor the overlap matrix we use the method of spectral decomposition. Generally, $N_\mathrm{block}$ will be small enough to diagonalize quickly. Singular or near-singular values are also easy to locate and remove. To compute the $\mathbf{B}_n$, we introduce a unitary matrix $\mathbf{U}$, constructed via the eigenvectors of $\mathbf{O}$, while $\mathbf{\Lambda}$ is the matrix of postive eigenvalues of $\mathbf{O}$; then $\mathbf{O}=\mathbf{U\Lambda U^T}$. From this we we construct
\begin{equation}
\begin{aligned}
\mathbf{B}_n =\sqrt{ \mathbf{\Lambda}\mathbf{U}^T}.
\end{aligned}
\end{equation}
The square root is  trivial as $\Lambda$ is diagonal with positive-definite elements.
The next Lanczos block is
\begin{equation}
\begin{aligned}
\mathbf{V}_{n+1} = \mathbf{W}_n^{'}\mathbf{B}_n^{-1} = \mathbf{W}_n^{'}\mathbf{U}\mathbf{\Lambda}^{-\frac{1}{2}}.
\end{aligned}
\end{equation}
The representation of the Hamiltonian in the new basis is 
\begin{equation}
H^{k} = 
\begin{pmatrix}
\mathbf{A}_{1} & \mathbf{B}_{1} & &  0 \\
\mathbf{B}_{1} & \mathbf{A}_{2} &\ddots &\\
  &\ddots   & \ddots & \mathbf{B}_k  \\
0 & & \mathbf{B}_{k-1} & \mathbf{A}_{k} 
\end{pmatrix},
\end{equation}
which is solved using conventional Lapack routines. The load balancing and memory locality improvements within our {\tt BIGSTICK} CI code show block Lanczos to run nearly twice as fast as single-vector Lanczos. However, one can further exploit  block Lanczos via bootstrapping. 

We `bootstrap' block Lanczos  by selecting vectors in the pivot block to be approximations of the lowest eigenvectors, for example by generating them in a much smaller 
model space and projecting into the current working space.
In  Section \ref{sec:level3}, we show how bootstrapped block Lanczos time-to-solution depends upon the dimension of the truncation used to construct the pivot block. 

While we found this worked extremely well and led to 
a dramatic reduction in number of iterations, there is an important caution: there must be a nontrivial overlap 
between the final, converged eigenvector and the pivot block. A contrary mismatch can happen, for example,  
 if 
the pivot block has an vector with quantum numbers such as angular momentum $J$ not found in the final converged 
eigenvectors, because of level crossing as the model space is increased; we give examples below. 
In nuclear physics, states which 
are mostly or completely orthogonal to a model space are referred to as `intruder' states.
The analogous existence of intruders in  the final converged eigenstates 
leads to needing an increased number of iterations. This is easily explained. One can understand the Lanczos 
algorithm as increasing the amplitude of low-lying states with each iteration, but when the initial amplitude 
is tiny, or must be generated through round-off error, it takes longer for such a state to converge.

\section{\label{sec:level3}Results}

In the previous section we outlined the conventional single-vector Lanczos, block Lanczos, and bootstrapped block Lanczos eigensolvers for the large-dimension eigenvalue problem. 
In this section we discuss the convergence and performance properties of the three methods used to calculate the eigenstates of two distinct nuclear systems. 

For the nonexpert, the only important difference between the two 
systems is the efficiency of our code. In all of our variations on the 
Lanczos algorithm, the fundamental step is the multiplication of a 
matrix on a vector or block of vectors. For us, the matrix is the 
nuclear many-body Hamiltonian, which is a very sparse but very large 
matrix, so large that the number of nonzero matrix elements can be a 
challenge to store.  Instead, we use a factorization algorithm to 
reconstruct the matrix elements on-the-fly. The memory required to 
store the arrays (called `jump' arrays) from which one reconstructs the matrix is at least one or two orders of magnitude smaller than 
the memory that would be required to store the resulting nonzero matrix elements. (To understand this reduction in memory: the physical 
Hamiltonian encodes interactions between two or three particles at a 
time; the reduction occurs by looping over spectator particles.) 
This factorization is most efficient, both in the reduction of memory 
costs and in the time to reconstruct the matrix elements, for 
full configuration spaces, that is, given a finite set of single-particle 
orbitals, the space of all possible configurations (subject to 
restrictions arising out of conservation laws). Our code can also 
truncate the many-body space through additional restrictions on 
allowed configurations. For these truncated spaces the factorization is less efficient.
Although some more details can be found in the Appendix, in brief: 
the factorization  works by breaking up data into blocks, 
and the efficiency of the code depends upon the length of the blocks.
The metaphor we use is to consider a rectangle: the stored data is 
the perimeter of a rectangle, while the reconstructed data is the area 
of the rectangle. The larger the rectangle, the greater the ratio of the 
area to perimeter and the more efficient.
For full configuration spaces one has fewer, larger blocks,  
truncated configurations have more, smaller blocks.

Thus, for our test cases, we consider a full configuration space and a 
highly truncated space. The first case is more efficient in reconstructing the matrix elements than the second. 
(Additionally, although not germane to our main focus here, in the first case the main memory burden is in storing the Lanczos vectors; in the 
second case, because the factorization is less efficient, the 
memory burden is actually in storing the `jump' arrays used to reconstruct the matrix elements.)
We measure the efficiency by computing the time per operation, that is, in each application of a single matrix elements.
Below in Tables~\ref{tab:53MnHvec_iter} and \ref{tab:7Betimeop} one can see that in the second case vector Lanczos requires nearly a third more time per operation than the first case.  In both cases block Lanczos reduces the 
time per operation to roughly the same quantity; there is a greater speed-up for the second case because of the 
greater starting inefficiency.

\begin{figure}[h]
	 \includegraphics[width=0.7\linewidth]{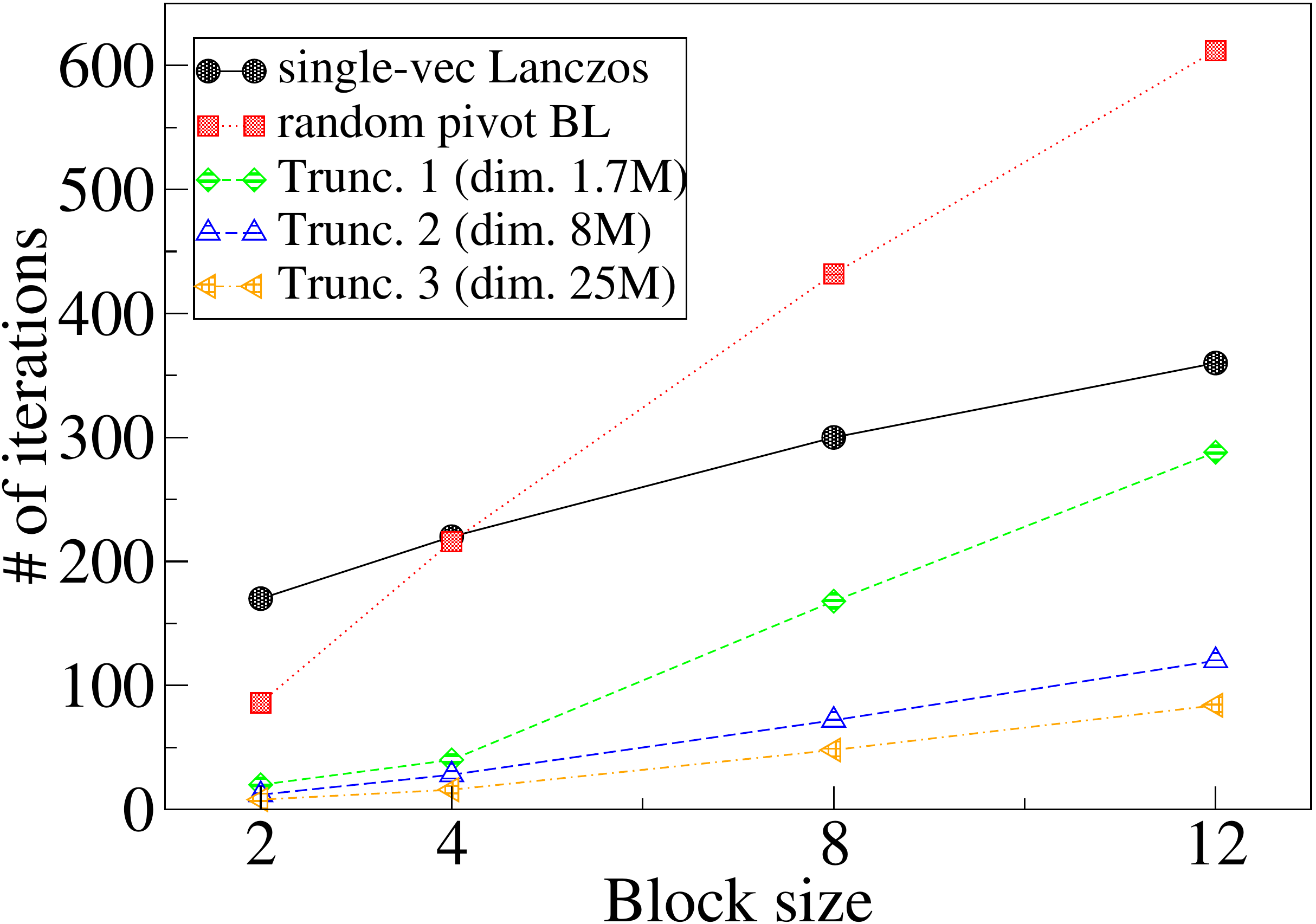} 
	 \caption{Number of vector-equivalent iterations (that is, the number of block iterations times the block size)  needed to reach convergence with block Lanczos using a randomly constructed pivot, and using pivot blocks constructed in three different truncated model spaces, for FCI calculations of $^{53}$Mn. 
	 The number of converged eigenstates is the block size.  Each truncation is labeled by its M-scheme basis dimension. Single-vector Lanczos results are shown for comparison.}
	\label{fig:53Mniters}
\end{figure}

\subsection{Case 1: $^{53}$Mn}

Our first test case is a full configuration calculation:
 $^{53}$Mn in the $pf$ valence space, consisting of the 
$1p_{1/2,3/2}$ and  $0f_{5/2,7/2}$ orbitals and 
 assuming a frozen $^{40}$Ca core, using the GX1A empirical interaction~\cite{HonmaOtsuka2005}, which 
 reproduces experimental data well. 
 The full M-scheme model space has a dimension of 141 million. 
(Storage of two single-precision Lanczos vectors, the minimum needed, 
is just over 1 Gb, although in practice one also needs to keep all the Lanczos vectors for reorthogonalization.  By contrast, storage of the jump arrays needed for 
reconstructing the matrix elements is 0.7 Gb; to store all the 
reconstructed nonzero matrix elements would require over 800 Gb.) 

In addition to random pivots, we 
 constructed three bootstrapped pivot blocks each from a truncated model space.  Here we truncated by restricting the number $W$ of particles excited out of the $0f_{7/2}$ orbital, which empirically has the 
 lowest energy. 
 Full configuration would be $W=13$, allowing all 13 nucleons to be 
 excited. Our truncations for generating the bootstrapping pivots 
 are  $(W=4)$ with a dimension of 1.65 million;  $(W=5)$, dimension of 7.95 million; and  $(W=6)$, with a dimension of 25.5 million. We 
 refer to these as Truncations 1, 2, and 3, respectively.
 Computations were performed on a computer outfitted with $32$ Intel(R) Xeon(R) CPU cores. The work was distributed over $32$ OpenMP threads, 
 though for timing purposes we extract the time for each thread, i.e., 
 we multiplied our raw timing numbers by 32.

Fig.~\ref{fig:53Mniters} shows the total number of Lanczos iterations 
to reach convergence (defined as $< 1$ keV rms change in energies) starting from the pivots constructed in the three different model spaces, for different block sizes. FCI calculations using single-vector Lanczos and conventional block Lanczos are shown for comparison. We chose the target number of eigenstates to converge to be 
equal to the block size, so naturally the number of iterations increases 
with block size/number of converged eigenstates.

\begin{figure}[b]
	 \includegraphics[width=0.7\linewidth]{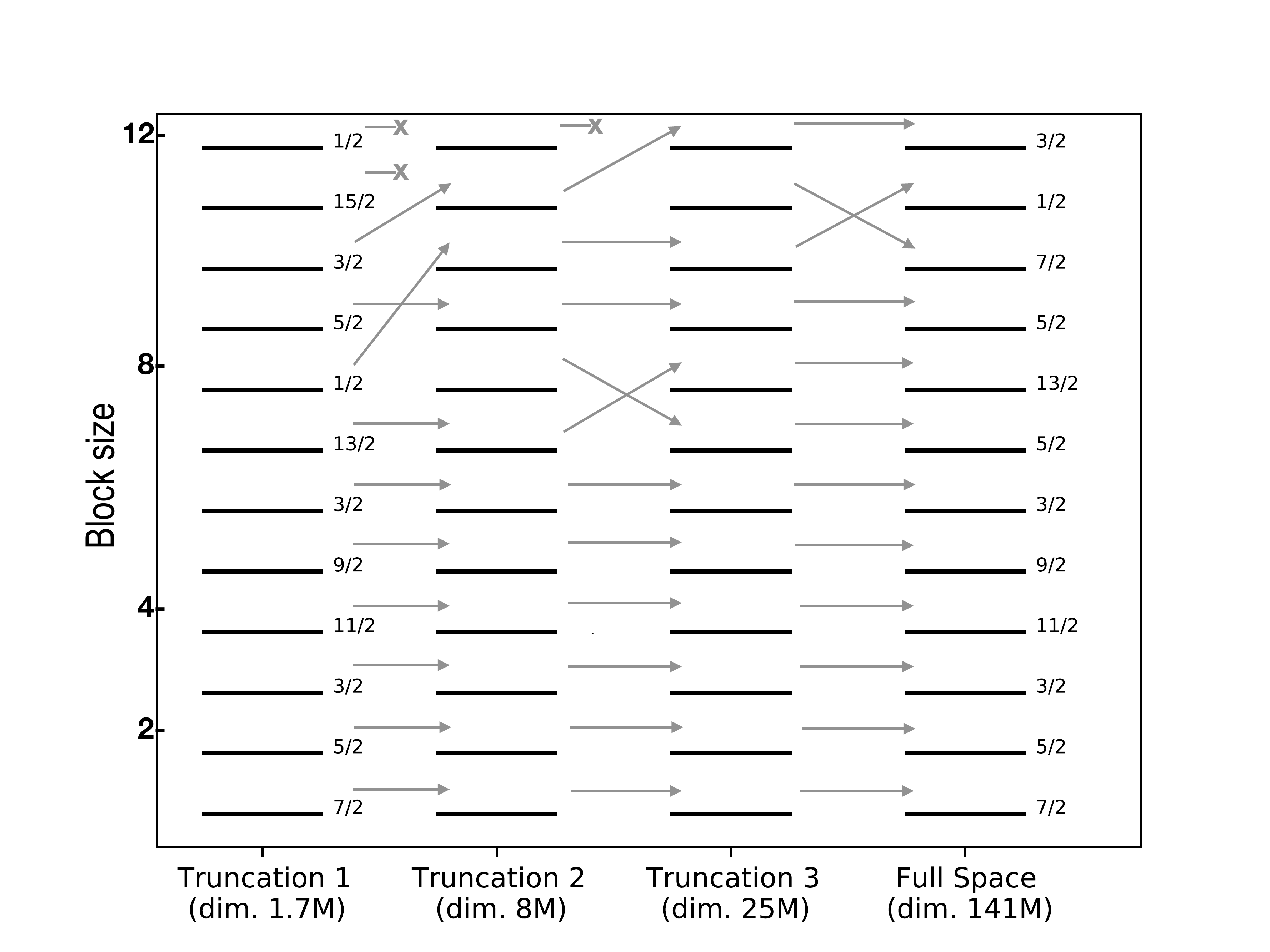} 
	 \caption{Excitation spectra of $^{53}$Mn computed in each truncated model space as shown in Fig.\ref{fig:53Mniters}, and the full model space. Each level within the spectra corresponds to an eigenstate labeled vertically by the state's total angular momentum $J$. Arrows connect eigenstates with strong overlap between model spaces leading up to the desired full model space. Arrows terminating with an $\times$ indicate states with no overlap with the other spectra states.}
	\label{fig:Mn53levels}
\end{figure}

To compare single-vector and block Lanczos and across different 
block sizes, we convert all calculations to  vector-equivalent  
iterations, by multiplying the number of block iterations by the block size. So, for example, 10 iterations of block Lanczos with a block size of 4 would have 40 vector-equivalent iterations. 
For block sizes two and four, each truncation of bootstrapped pivot caused a comparable reduction in iteration count. However,  for block sizes greater than four, the truncation $1$ pivot had more intruder states (see discussion below) leading to a rise in number of 
vector-equivalent 
iterations needed.
The pivots constructed using eigenvectors from the other two truncated model spaces provided consistently iteration counts.  For comparison, FCI calculations using block Lanczos with a block size of 12 starting from  a randomly constructed pivot took $612$ vector-equivalent iterations to reach convergence, 
while single-vector Lanczos took $360$ vector-equivalent iterations to converge 12 states, 
while starting from a pivot constructed from the ($W=6$) truncation, only $84$ vector-equivalent iterations.

The differences in iteration count can be understood through 
the concept of `intruders', as illustrated in   Fig.~\ref{fig:Mn53levels}. Here we define an intruder as a 
state in one set (either the pivot block or the final converged 
eigenstates) which has a very small overlap with all states in the 
final set. Very small overlaps slow convergence. In particular, 
because converged states have total angular 
momentum $J$ as a good quantum number, if any state in the pivot 
block has a $J$ not found in the final set of converged states, it 
is orthogonal to the converged states and does not contribute;
conversely, if a converged state has a $J$ not found in the pivot 
block it too will converge very slowly. (We have experimented using 
only partially converged states for our pivot, but did not obtain 
conclusive results.)

\begin{figure}[h]
\centering
	\includegraphics[width=0.7\linewidth]{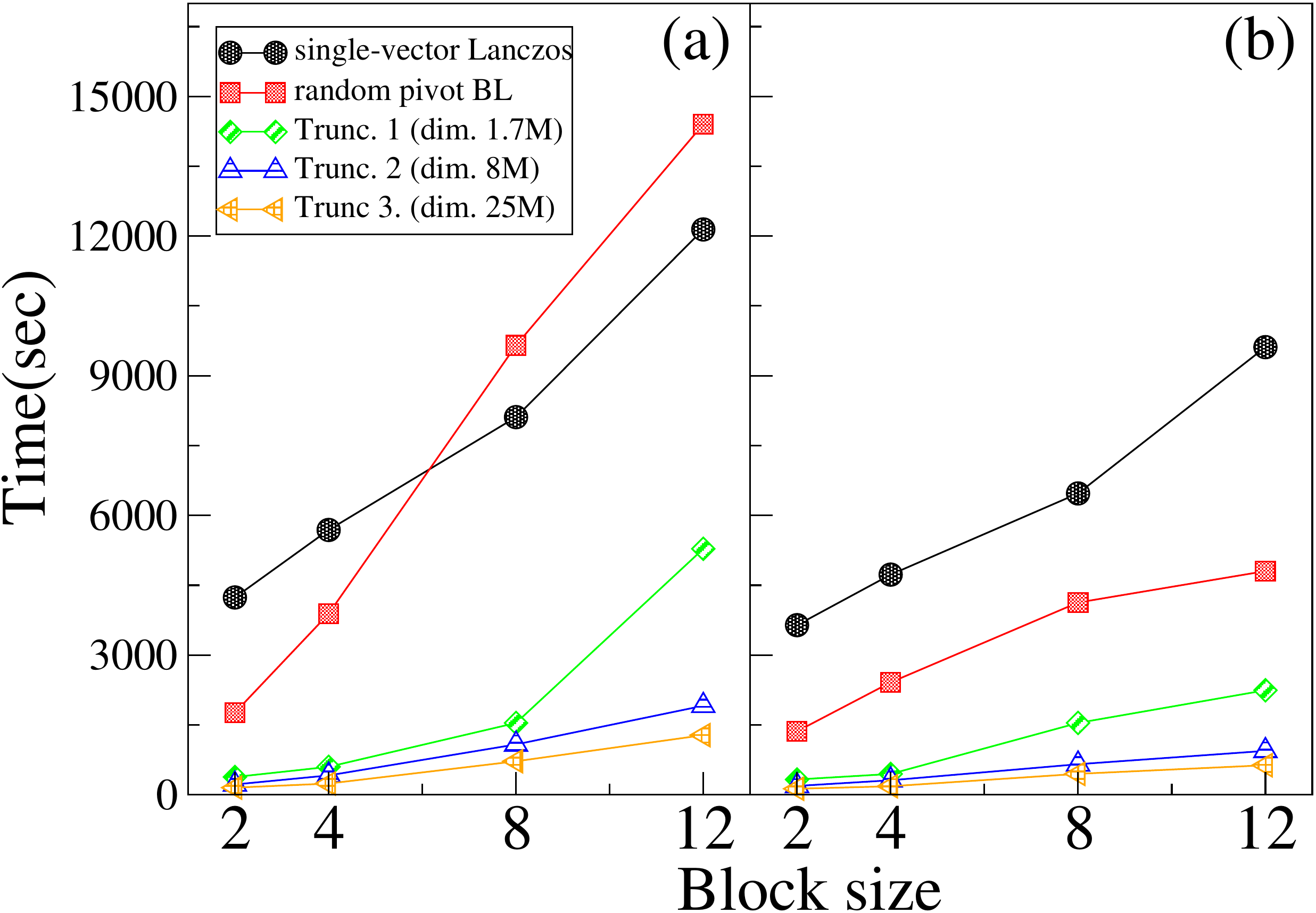}
		\caption{Panel (a): Total time-to-solution for each scenario in Fig.~\ref{fig:53Mniters}; Panel (b): Total equivalent matrix-vector multiplication time.}
	\label{fig:53MnlancHvec}
\end{figure}

Truncation $1$ has multiple higher-lying intruder states ( $J=\nicefrac{15}{2}$, $J=\nicefrac{1}{2}$ ) orthogonal to the full model space spectra as seen from the terminating arrows -$\times$. We also observe significant level crossings as one goes from the smallest truncated space to the full space. This explains the increasing in  the total Lanczos time  in Fig.~\ref{fig:53MnlancHvec}(a) for block size $12$ and not block size $8$. A single orthogonal intruder $J=\nicefrac{1}{2}$ state is also seen in truncation $2$. However, there are no intruding orthogonal states seen in truncation $3$ up to block size $12$. 
Our conclusion, based upon this and many other trials, is that the danger of using a pivot block from 
the most severely truncated spaces is not from the truncation itself, 
but from a reordering of the final spectrum leading to `intruder' states.

The number of iterations to convergence, while important, is not 
the only contribution to a speed-up using block Lanczos. 
Fig.~\ref{fig:53MnlancHvec}(a) shows the time-to-solution for each method.  Reducing the number of iterations can reduce the 
time cost of reorthogonalization, which increases quadratically 
with the number of iterations. This can be seen by comparing 
Fig.~\ref{fig:53MnlancHvec}(a) to Fig.~\ref{fig:53MnlancHvec}(b), 
which shows the matrix multiplication performance.

Even without the savings in iterations from bootstrapping, 
block Lanczos leads to a significantly faster algorithm just in 
matrix-matrix multiplication.
Table~\ref{tab:53MnHvec_iter} shows the time per operation 
(reconstructing and applying a single nonzero matrix element).
Note that while our results were obtained using 32 OpenMP threads, we show the 
time per thread, so that the time per operation is independent 
of the number of threads; although we don't show it, for large 
FCI calculations such as this our core routines scale well with the 
number of threads. 
Here the power of block Lanczos is clear: by 
improving data locality and amortizing the cost of reconstructing 
the matrix elements, 
we observe up to  $3 \times$ reduction in 
equivalent matrix-vector multiplication time.

\begin{table}[h]
\begin{center}
\begin{tabular} { | c | c |  }
\hline
 block size & time per operation (ns)/(op) \\
 \hline
 1 & 3.3 \\
 \hline
2&  	2.5\\
 \hline
4&  1.7\\
 \hline
8&  	1.5\\
 \hline
12& 	1.2\\
 \hline

\end{tabular}
\end{center}
\caption{Time to construct and apply a single matrix element (operation) 
as a function of block size, FCI calculations of $^{53}$Mn. 
A block size of 1 is the time for single vector Lanczos.}
\label{tab:53MnHvec_iter}
\end{table}

To conclude our first case, let us summarize  the reduction in  time-to-solution spent performing Lanczos in the bootstrapped block size$=12$ calculations. Conventional block Lanczos with random pivot vectors took $14 \times 10^3$  s to reach convergence for a block size $12$. Single-vector Lanczos converged $12$ eigenpairs in $12 \times 10^3 $s. With a $W=6$ bootstrapped pivot, however, block Lanczos took only $1.28 \times 10^3$s, a nearly 9$\times$ speedup over single-vector Lanczos, and an approximately 11$\times$ speedup over conventional block Lanczos. Our results support bootstrapping as a competitive method of improving the conventional block Lanczos approach.

\subsection{Case 2: $^7$Be}

Here we turn to a different case, a no-core shell model (NCSM) 
calculation of $^7$Be, which differs from the FCI calculation of
$^{53}$Mn in several  ways. 
Most important for the work reported here, the space is not based simply on  available orbitals, but rather on the excitations of a 
non-interacting harmonic oscillator, the so-called $N_\mathrm{max}$ 
or $N\hbar \Omega$ truncation, which is easily handled by our code. 
For more details see
the Appendix and the 
literature~\cite{navratil2000large,navratil2009recent,barrett2013ab}. 
The key consequences are that the reconstruction of matrix 
elements on-the-fly is less efficient both in memory and time 
requirements.

We choose as our target space the  $N_\mathrm{max} =8$ truncation, 
which is not fully converged but is suitable for detailed investigations of Lanczos schemes.
The $^7$Be M-scheme dimension is $6.1$ million, for which the storage of two single-precision Lanczos vectors requires 
only 49 Mb, while storage of the jump arrays requires 2.8 Gb; storage 
of the nonzero matrix elements would, by comparison, require 35 Gb. 
This differs significantly from the FCI calculation where the 
storage of Lanczos vectors competes with and outstrips storage of 
the jump arrays.
Although not crucial to our results, we use a  two-body interaction derived from an \textit{ab initio} N3LO chiral effective field theory~\cite{MACHLEIDT20111}.   Because calculations of such truncated 
spaces do not scale as well with the number of OpenMP threads, we
worked with only 8 threads.

To generate the 
approximate eigenvectors used to construct bootstrapped pivot blocks,
we worked in three different truncated $M$-scheme model spaces, each restricted to a different maximum harmonic oscillator excitation $N_\mathrm{max}$ ($N_\mathrm{max}=2$, $N_\mathrm{max}=4$, $N_\mathrm{max}=6$). The dimension of the $N_\mathrm{max}=2$  basis is $1961$, $N_{max}=4$ is $49$ thousand and the third, $N_\mathrm{max}=6$, is $660$ thousand. We refer these spaces  as truncation 1-3, respectively.

\begin{figure}[h]
	 \includegraphics[width=0.7\linewidth]{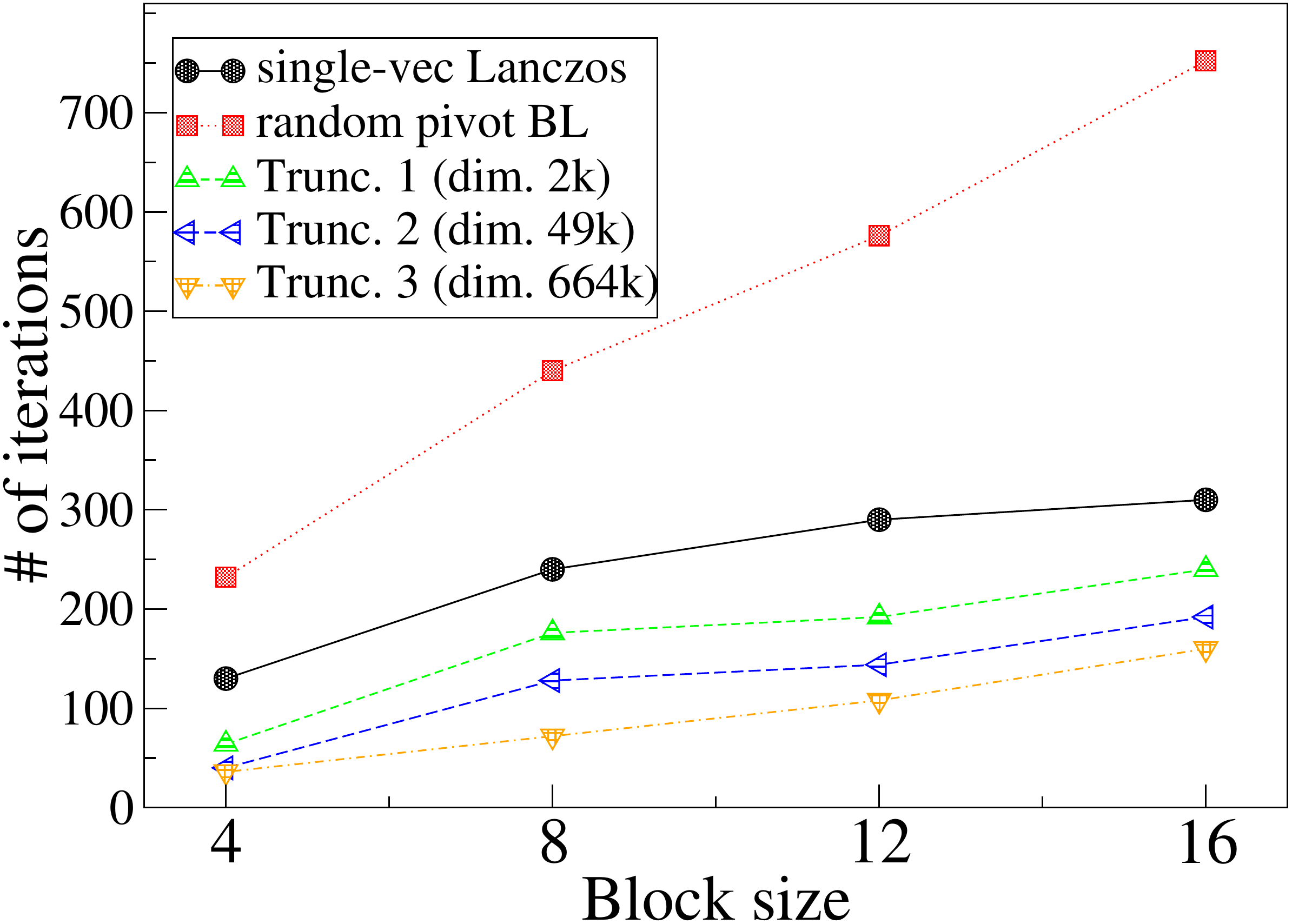} 
	 \caption{The same as Fig.~\ref{fig:53Mniters}, but for $N_\mathrm{max}=8$ calculations of $^{7}$Be in the no-core shell 
	 model.}
	 %
	\label{fig:be7iters}
\end{figure}

Fig.~\ref{fig:be7iters} shows the number of Lanczos iterations needed for each method to converge a number of eigenstates equal to the block size. 

Bootstrapped block Lanczos, using the pivot from truncation $1$ converges in about half the total number of vector-equivalent iterations as block Lanczos with a random pivot. Using larger model spaces to compute the pivot block produced minimal gains. As we found with 
our FCI $^{53}$Mn case, better approximations to eigenvectors as 
pivots do not translate into significant reductions in the 
number of iterations. 

More important is the elimination of 
`intruders' from the pivot block; avoiding intruders is the main advantage of using a larger space to generate the pivot.  In the case of $^7$Be, there 
were fewer level crossings and thus fewer intruders.
In Fig.~\ref{fig:be7iters}, the number of iterations to converge $8$ eigenpairs (block size $8$) is relatively close to that needed for $12$
eigenpairs (block size $=12$) for each pivot truncation. This is because the first twelve pivot vectors calculated in the truncated spaces having significant overlap with twelve lowest eigenvectors in the target $N_\mathrm{max} =8$ space. While performing the bootstrapped block methods, intruder states, that is states outside the desired model space enter for block sizes larger than $12$. More specifically, the Krylov subspace converges to a spectra where some eigenstates lack physical meaning. A rise in iteration count can be seen in for block sizes $=16$ as some of the states within the initial pivot block beyond block size$=12$ are intruders.

To converge 16 eigenvectors, block Lanczos with a random pivot took $752$ vector-equivalent iterations. Bootrapping using pivot vectors from truncation $1$ required only $240$ iterations, a significant improvement.
Although we do not show it, we also compared the convergence for 
32 eigenvectors using a block size of 32. Block Lanczos with 
random pivot vectors  took $1760$ vector-equivalent iterations, while bootstrapping from a truncation $1$ bootstrapped calculation 
took $1152$ vector-equivalent iterations, bootstrapping from truncation $2$  took $864$ vector-equivalent iterations, and bootstrapping from truncation $3$ approximate eigenvectors took $544$ vector-equivalent iterations. 

\begin{figure}[h]
\centering

	\includegraphics[width=0.7\linewidth]{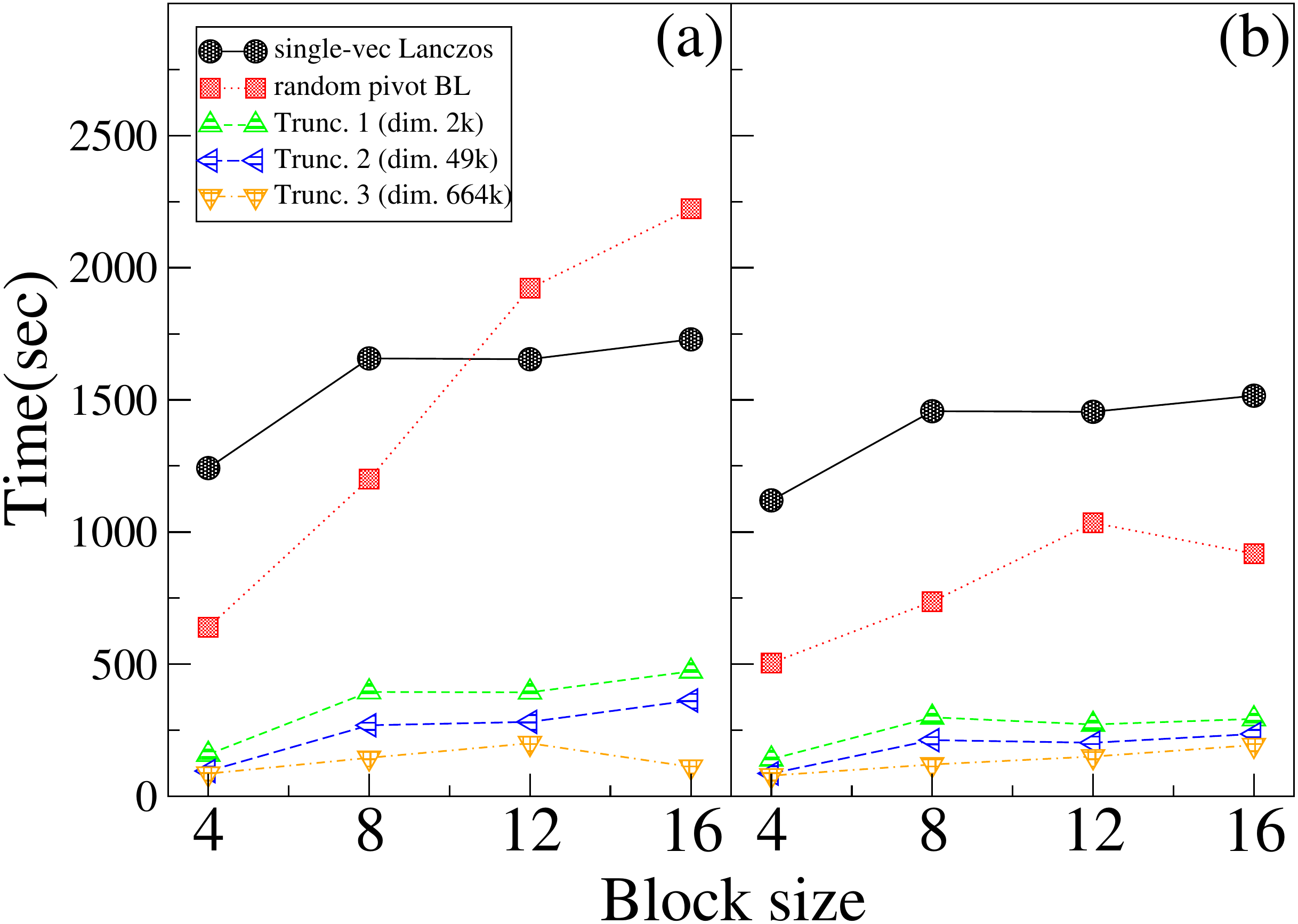}
		\caption{  ${}^{7}$Be, computed at $N_\mathrm{max} =8$ with a single-particle harmonic oscillator basis frequency $\hbar \omega = 20$ MeV, using a two-body chiral force at N3LO. Panel (a):Total Elapsed Lanczos algorithm time for each pivot truncation as seen in Fig.~\ref{fig:be7iters}. Panel (b):Total Elapsed Hamiltonian-vector multiplication time for each truncation of pivot as seen in Fig.~\ref{fig:be7iters}. Single-vector and conventional block Lanczos with random pivot vectors are also shown for comparison. Calculations using block Lanczos with a bootstrapped pivot are labeled by the dimension of the truncated model space.}
	\label{fig:7belancHvec}
\end{figure}

As with our full configuration calculations of $^{53}$Mn, 
the reduction of vector-equivalent iterations pays dividends in 
the significantly improved time-to-solution shown in 
 Fig.~\ref{fig:7belancHvec}(a). (We should note here that the 
 time spent outside the Lanczos algorithm, in constructing the basis 
 and the jump arrays for reconstructing the matrix elements, etc., 
is negligible compared to the time spent in the actual Lanczos algorithm.)

  The advantage of the matrix-matrix multiplication of the block method becomes clear in Fig.~\ref{fig:7belancHvec}(b);  the time spent in Hamiltonian-block multiplication for conventional block Lanczos is nearly half of single-vector Lanczos for each block size. However, Fig.~\ref{fig:7belancHvec}(a) also highlights a weakness of the block Lanczos approach. That is, the time spent in reorthogonalization grows linearly with block dimension. This is strong motivation for the  usage of a bootstrapped thick-restart approach to these calculations when a large number of low-lying states is desired. 
  (We have recently added this capability to our code, but 
  choose to focus on non-restarted calculations here.)

Our bootstrapped Lanczos results demonstrate strong improvements are possible  even without restarting Lanczos. Using pivot vectors constructed in the truncation $1$ model space, which is less than half a percent of the full $N_\mathrm{max}=8$ dimension, resulted in a factor of two to three speedup. For block size of $16$, single-vector Lanczos had a total Hamiltonian-vector multiplication time of $1500$s, conventional block Lanczos with random pivot, $929$s and bootstrapped block Lanczos using truncation $1$ eigenvectors took $290$s. We emphasize the first truncation here based on its trivial computational cost relative to observed benefit in speedup.

As a further study, we also tried $32$ eigenvectors block size of $32$. Matrix vector times for single-vector Lanczos took $3600$s, random pivot block Lanczos took $1800$s and bootstrapped block Lanczos using the first truncation pivot, $1200$s.


Interestingly for truncations $1$ and $2$, the bootstrapped Hamiltonian-vector time for block size of $8$ is near sizes $12$ and $16$ as shown in Fig.~\ref{fig:7belancHvec}(b). This additional speedup seemed to be related, firstly, to improved load balancing producing faster operation times as seen in Table~\ref{tab:7Betimeop} despite a larger block size. 
Another contributing factor to the speedup is the larger block sizes have fewer intruder states for the chosen truncations.
Table~\ref{tab:7Betimeop} compares the total Hamiltonian-vector time in nanoseconds per operation per iteration using $8$ OpenMP threads of single-vector Lanczos and random pivot block Lanczos. (We normalize 
 by multiplying the raw timing data by 
the number of threads.) Note that despite the very different 
dimensionalities and truncation schemes, for large block sizes 
we achieve very nearly the same time per operation as for the 
FCI calculations of $^{53}$Mn.


\begin{table}[h]
\begin{center}
\begin{tabular} { | c | c | }
\hline
block size & time per operation (ns)/(op)\\
\hline
1 & 4.3 \\
\hline
4& 	2.0\\
\hline
8&  	1.5\\
\hline
12& 	1.6\\
\hline
16& 	1.1\\

\hline
\end{tabular}
\end{center}
\caption{$^7$Be Hamiltonian-vector algorithm time in nanoseconds per operation for different quantities of desired converged eigenvectors. Calculation were performed using $8$ OpenMP threads.}
\label{tab:7Betimeop}
\end{table}

\section{\label{sec:level4}Conclusions and acknowledgements}

The Lanczos algorithm is widely used, in part because it is 
well-suited to many physical problems where one wants to 
find extremal eigenpairs of very large, very sparse matrices. 
The sparsity leads to loss of 
data locality, while very large matrices may be addressed 
with sophisticated on-the-fly reconstruction of the 
matrix elements, which reduces the memory footprint but 
costs additional time.  Block Lanczos can address both 
those issues, by improving data locality and amortizing the 
cost of reconstructing matrix elements. 
Indeed, in our experiments, implemented in a configuration-interaction nuclear structure code {\tt BIGSTICK}, 
and tested in two very different computational scenarios, 
we found block Lanczos significantly sped up the intrinsic speed of sparse matrix 
multiplication, as measured in  time-per-operation, that is, the time to reconstruct and apply
a matrix element, by a factor of 3 or more. 

In practice, however, 
one often finds that block Lanczos starting from a random pivot requires more vector-equivalent
iterations than standard single vector Lanczos. 
To address this, we carefully studied bootstrapping block Lanczos,
by using an approximate pivot block constructed from 
eigenpairs found in a dramatically truncated space. 
We implemented bootstrapped block Lanczos in our 
configuration-interaction nuclear structure code {\tt BIGSTICK}, 
and tested two very different computational scenarios. 
Bootstrapped block Lanczos works very well and significantly reduces the 
number of iterations required and thus the time-to-solution. 
The one caveat is that all the final converged eigenvectors 
need to have a nontrivial overlap with the initial pivot block;
if there are `intruders,' that is, final converged eigenvectors 
with no or very tiny overlap--for example, if an eigenvector 
has a quantum number not found in the pivot--the time required to 
converge that eigenvector is dramatically more. (In general, even if a 
final eigenvector has zero overlap with the initial pivot, numerical noise 
will eventually cause that eigenvector to appear, albeit at a cost of 
additional iterations.)  Pivots constructed from approximate eigenpairs in a severely truncated space nonetheless provide good enough 
starting vectors for bootstrapping; the main utility of constructing pivots from 
less truncated spaces is to avoid the problem of `intruders.'  
Because of the speed-up of time-to-solution, bootstrapped block Lanczos is a
very attractive alternative for large problems that require significant computer time.

This material is based upon work supported by the U.S. Department of Energy, Office of Science, Office of Nuclear Physics, under Award Number  DE-FG02-03ER41272.

\appendix

\section{The configuration-interaction context}

\label{CIcontext}

Although our conclusions can be appreciated independent of the specific science 
problem, we here provide some details on the context of our calculations, 
the configuration-interaction approach to the quantum many-body problem, and 
in particular wave functions for atomic nuclei. 

Configuration interaction (CI) is a popular method of constructing the basis for the nuclear many-body wave function~\cite{doi:10.1080/002689798168303}.
Here the many-body wave function is expanded in some convenient orthonormal basis $|\alpha\rangle$: 
\begin{equation}
\begin{aligned}
| \Psi \rangle = \sum_\alpha c_\alpha | \alpha \rangle.
\end{aligned}
\end{equation}

There are numerous benefits to CI; notably, it allows for arbitrary choice of single-particle basis, works well for shell-model calculations with no dependence on the form of two-body interaction in the creation of the Hamiltonian matrix, and one 
can construct low-lying excited states nearly as easily as the ground state. 
The major drawback of the CI method is that the dimension of the basis  grows exponentially with the number of nucleons.


In nuclear physics CI calculations~\cite{BG77,br88,ca05} it is common, though not required, to use a harmonic oscillator single particle basis.
Nuclear CI calculations can be divided into two main categories, \textit{ab initio} or also called \textit{no-core shell model} (NCSM) and \textit{empirical} or phenomenological. 
In empirical many-body spaces, one has an inert core with active orbits occupied up to a designated maximum. The active space or \textit{valence} space is a fixed set of orbits.  Phenomenological interactions are adjusted to the many-body spectra within the given shell.  No-core shell model many-body bases allow all orbits to be 
active, specifying a model space by a maximum number of excitation quanta above the lowest configuration $N_\mathrm{max}$~\cite{navratil2000large,barrett2013ab}. 
\textit{Ab initio}/NCSM interactions are derived from two- and few-body data. 

We specifically choose the basis to be a linear combination of antisymmetrized products of single particle wave functions or \textit{Slater determinants} representing the ways particles can occupy active orbitals. Each Slater determinant has the same fixed $z$-component of angular momentum, forming an M-scheme basis\cite{ca05}. There are many well known CI shell model codes which use a M-scheme basis, ANTOINE~\cite{caurier:in2p3-00003301},  MFDn~\cite{MFDN}, {\tt BIGSTICK}~\cite{johnson2018bigstick},  and KSHELL~\cite{shimizu2013nuclear}, all of which except MFDn are publicly available. M-scheme bases are ideal for bit occupation representation, and efficient computation of matrix elements as we do not explicitly use the Slater determinants. Instead we use the \textit{occupation representation} of the Slater determinants using fermion creation and annihilation operators. In the CI basis, states have a discrete and fixed number of fermions. Such a basis in principle can be used to describe any nucleus although computational practicalities 
limit the choice of both nuclides and model spaces. 

While $M$-scheme dimensions are large, the matrices are very sparse, on the 
order of only a few  elements in a million being nonzero. Nonetheless, 
for large dimensions the total number of matrix elements to be stored can be 
huge.  (Other bases, 
such as fixed total angular momentum or $J$-scheme bases, or those based upon 
other groups, have smaller dimensions but are denser, with larger number of 
nonzero matrix elements~\cite{dytrych2016efficacy}.)  Here a physical picture 
helps: the Hamiltonian 
matrix represents interactions between two or possibly three nucleons, with the 
rest as spectators, a picture which allows us to understand the sparsity of the 
matrices. Furthermore, the idea of spectators allows one to factorize  the 
Hamiltonian~\cite{caurier:in2p3-00003301,Johnson2013}, which reduces the memory 
requirements by up to two orders of magnitude. In that case, however, one must 
reconstruct the matrix elements ``on-the-fly'' which, no matter how 
efficient, incur a time cost.  The block Lanczos algorithm described above 
amortizes that cost, contributing to the speed-up of factorized or ``on-the-fly'' 
codes.

In {\tt BIGSTICK}, the data is organized by Abelian quantum numbers such as $M/J_z$ and 
parity~\cite{Johnson2013,johnson2018bigstick}. This makes it easy to aggregate information in blocks. For example, 
a given calculation always has fixed total $M$, which is the sum of the 
values from the protons and the neutrons, $M=M_p + M_n$.  Thus we automatically know that 
proton configurations with a given $M_p$ must connect up with neutron $M_n= M-M_p$.
This scheme allows us to `factorize' data; the metaphor we use is that of a 
rectangle, with the perimeter being the stored data and the area the reconstructed data. 
So, continuing the example of the basis, the basis states are represented by 
rectangles with edges representing proton configurations with $M_p$ and neutron 
configurations with $M_n$ and the interior the combined basis states with total 
$M=M_p+ M_n$.  

To truncate the space, we add a nonphysical quantum number, $W$. This is an nonnegative 
integer which is assigned to each single particle orbital. Then we restrict to 
configurations with total $W \leq W_\mathrm{max}$~\cite{johnson2018bigstick}. While not infinitely flexible, 
it nonetheless encompasses many standard truncations, such as particle-hole expansions.
In particular, it encompasses the $N_\mathrm{max}$ or $N\hbar \Omega$ truncation we use. 
Here the $W$-value of any single-particle orbital is the principal quantum number $N$ of 
a spherically symmetric harmonic oscillator. These orbits are grouped into `major shells,' 
such as the $0s$, which has $W=N=0$, the $0p$ which has $N=1$, the $1s0d$, which has $N=2$, 
and so on. For any given nuclide, one computes the lowest possible configuration value of 
$W$ compatible with the Pauli exclusion principle for fermions. In the case of $^7$Be, 
the lowest configuration has two protons and two neutrons into the $0s$ shell, each with $W=0$ and the remaining three nucleons go into the $0p$ shell, each with $W=1$, so that the 
minimum total $W_\mathrm{min}=3$.  For a given $N_\mathrm{max}$, one allows excitations up to 
$W_\mathrm{max}=W_\mathrm{min}+N_\mathrm{max}$. So, for $N_\mathrm{max}=8$, one particle 
could be excited up 8 shells, or 2 particles up to 4 shells each, or 4 particles each up one shell, or 1 particle up to 5 
shells and 1 particle up to 3 shells, and so on. Because each major shell in the spherical 
harmonic oscillator is separated by $\hbar \Omega$ in energy, this means one allows up to 
$N_\mathrm{max} \hbar \Omega$ excitation energy in the non-interacting harmonic oscillator, 
which crudely approximates the effect of the nuclear mean field.

Nonetheless, adding the $W$ truncation
makes the code run less efficiently; again, one can understand this metaphorically, 
that the ratio of area to perimeter is largest for a few large rectangles compared to 
many small rectangles.

\bibliography{johnsonmaster}

\end{document}